\def\rfr#1{eq.(\ref{#1})}

\def\Rfr#1{Eq.(\ref{#1})}

\def\eqi{\begin{equation}}
\def\eqf{\end{equation}}
\def\eqia{\begin{eqnarray}}
\def\eqfa{\end{eqnarray}}
\def\rp#1#2{{#1\over#2}}
\def\ct#1{\cite{#1}}
\def\lb#1{\label{#1}}
\def\bm#1{{\mbox{\boldmath$#1$\unboldmath}}}

\def\cq{c^2}

\def\velocità{\rp{2C_2 e^{\rp{g(t-t_0)}{c}}\left[1+c^2 C_1 C_2^2 e^{\rp{2g(t-t_0)}{c}}\right]}{c\left[-1+\cq C_1 C_2^2 e^{\rp{2g(t-t_0)}{c}}\right]^2}}

\documentclass[11pt]{article}

\usepackage{amsmath,amsthm,amscd,amssymb}
\usepackage{latexsym}
\usepackage{graphics,graphicx}

\begin{document}

\noindent{\bf \LARGE{On the Clock Paradox in the case of circular
motion of the moving clock}}
\\
\\
\\
Lorenzo Iorio\\Dipartimento Interateneo di Fisica dell'
Universit${\rm \grave{a}}$ di Bari\\
INFN-Sezione di Bari
\\Via Amendola 173, 70126\\Bari, Italy
\\
\\

\begin{abstract}
In this paper we deal analytically with a version of the so called
clock paradox in which the moving clock performs a $circular$
motion of constant radius. The rest clock is denoted as (1), the
rotating clock is (2), the inertial frame in which (1) is at rest
and (2) moves is $I$ and, finally, the accelerated frame in which
(2) is at rest and (1) rotates is $A$. By using the General Theory
of Relativity in order to describe the motion of (1) as seen in
$A$ we will show the following features. I) A differential aging
between (1) and (2) occurs at their reunion and it $has\ an\
absolute\ character$, i.e. the proper time interval measured by a
given clock is the $same$ both in $I$ and in $A$. II) From a
quantitative point of view, the magnitude of the differential
aging between (1) and (2) does depend on the kind of rotational
motion performed by $A$. Indeed, if it is $uniform$ there is no
any tangential force in the direction of motion of (2) but only
$normal$ to it. In this case, the proper time interval reckoned by
(2) does depend only on its {\it constant} velocity $v=r\omega$.
On the contrary, if the rotational motion is $uniformly\
accelerated$, i.e. a constant force acts $tangentially\ along\
the\ direction\ of motion$, the proper time intervals $do\ depend$
on the angular acceleration $\alpha$. III) Finally, in regard to
the sign of the aging, the moving clock (2) measures always a
$shorter$ interval of proper time with respect to (1).
\end{abstract}

\section{Introduction}\label{intro}
\subsection{The impact of the acceleration on the proper time of a moving clock}
What is the influence of the acceleration experienced by a moving
particle on the proper time $\tau$ reckoned by a clock at rest
with respect to it? The widely accepted answer is that $\tau$ is
independent of {\it any} form of acceleration because it depends
$explicitly$ only on the speed $v$ of the accelerating particle
according to the well known formula \eqi
d\tau=dt\sqrt{1-\left[\rp{v(t)}{c}\right]^2},\lb{dilat}\eqf in
which $c$ is the speed of light and $t$ is the time measured by a
standard inertial clock. \Rfr{dilat}, which has been directly
tested by observing the decay rates of fast moving elementary
particles in the cosmic rays \ct{muonsealevel}, is assumed to be
valid also in in the case of an accelerated motion of the moving
particle according to the hypothesis of locality \ct{locality}. As
an experimental support of the independence of acceleration the
experience with the circling muons in the CERN storage ring
\ct{muoncern} is often proposed. In it the muons are continuously
kept in a $uniform\ circular$ motion of constant radius $r$ at an
angular speed $\omega$ under the action of a $constant$ magnetic
field. Their observed decay rates are in agreement with
\rfr{dilat}, where $v=\omega r$; since the muons are {\it
continuously accelerated} by the $centripetal$ Lorentz force, this
fact would prove the independence of $\tau$ from the acceleration.
Very often this experience is adduced to present the independence
of $\tau$ from {\it any kind} of acceleration as if it was {\it a
general feature}.

Incidentally, let us note that the circling muons scenario could
also be considered as a version of the clock paradox in which the
moving clock, denoted conventionally as (2), moves along a
$circular$ orbit \ct{moller} instead of going to--and--fro in a
$straight\ line$. The rest clock will be conventionally denoted as
(1)\footnote{Its proper time would be, in this case, related to
the measured time of flight of the flying muons, while the proper
time measured by (2) would be related to the lifetime of the muons
at rest.}.
\subsection{Aim of the paper}
In a recent paper \ct{lortwin} the `rectilinear' version of the
clock paradox has been treated by accounting for the finite value
of a constant force \bm F which accelerates and decelerates (2).
In that case, in which a force acts {\it along\ the\ direction\
of\ motion} of the travelling clock, it has been found that
$g=F/m$, where $m$ is the mass of (2), {\it does\ affect\ the\
magnitude} of the proper time intervals $\Delta\tau$ measured by
the clocks at their reunion. However, this fact does not alter the
qualitative feature of the differential aging of the two clocks in
the sense that, also for $g\neq 0$,
$\Delta\tau^{(2)}<\Delta\tau^{(1)}$.

In this paper we wish to investigate what happens to the proper
time intervals $\Delta\tau$ if (2) reunites with (1) by performing
a {\it uniformly\ accelerated\ circular} motion, so that a {\it
tangential\ acceleration\ along\ the\ direction\ of\ its\ motion}
is present as well: indeed,
we hypothesize  that the result of the circling muons at CERN
could be due to the fact that, in that case, the acceleration
experienced by the particles is always {\it normal\ to\ their\
motion}. Moreover, we will use the General Theory of Relativity
in order to calculate $\Delta\tau^{(1)}$ and $\Delta\tau^{(2)}$ in
the accelerated frame in which (2) is at rest. We will investigate
if, in this case, the differential aging between the two clocks
can be considered, as expected, absolute, i.e.
$\Delta\tau^{(2)}<\Delta\tau^{(1)}$ both from the point of view of
the inertial reference frame, denoted conventionally as $I$, in
which (1) is at rest and (2) moves, and from the point of view of
the accelerated frame, denoted as $A$, in which (2) is at rest and
(1) is seen circling. Indeed, the elapsed proper time is function
only of the observer's worldline and of its starting/ending
points: it is the Lorentzian length of the segment of worldline
delimited by given endpoints. The spacetime of an accelerated
frame does not present a real curvature as if a true gravitational
field was present. Whatever coordinates are used for flat
spacetime, it will always be flat. The coordinates used might
change the form of the metric (even making it nondiagonal), but
they cannot create curvature. The choice of frame will also change
the coordinate expression of geodesics (e.g., the worldline of
clock (1)), but it will not change its geometrical properties,
including its proper-time length. Thus, the worldline will curve
in the sense that its spatial coordinates are not constant in the
accelerated frame, but it will still be straight in the sense that
it is inertial (and therefore an extremum of proper-time length).
\section{The point of view of the inertial observer}\lb{secuno}
Let us consider an inertial frame $I$ in which the usual cartesian
coordinates $\{X,Y,Z,T\}$ are used so that
\eqi(ds)^2=(cdT)^2-(dX)^2-(dY)^2-(dZ)^2\lb{minko}\eqf and a clock,
denoted as (1), is at rest. Another clock, denoted conventionally
as (2), moves nonuniformly along a circular path of constant
radius $r$ so to encounter (1) on its trajectory: let us assume
the center of the rotation of (2) as the origin of $I$ and
$\omega=\omega_0+\alpha T$ where $\alpha$ is a constant angular
acceleration. Let us consider a rotating frame in which (2) is at
rest: it will be denoted as $A$. According to Newtonian mechanics,
if (2) has to be at rest in $A$ {\it both centripetal and
tangential forces} must act on it. A real physical scenario in
which a particle moves around a circle of (nearly) constant radius
and its speed is continuously increased occurs, e.g., in the
betatron. Indeed, in it a spatially nonuniform, time-varying
magnetic field induces a rotational electric field along whose
lines of force an elementary electrically charged particle is kept
in uniformly accelerated circular motion \ct{feyn}.

Let us denote $\bar{T}\equiv\Delta\tau_I^{(1)}$ the time interval,
measured by (1), during which (2) describes an angular interval of
$2\pi$ starting when $T=0$. The proper time interval measured by
(2) is given by \rfr{dilat}.
In this case \eqi\Delta\tau_I^{(2)}=\int_0^{\bar{T}}
\sqrt{1-\left[\rp{r(\omega_0+\alpha
T^{'})}{c}\right]^2}dT^{'}.\lb{tempo}\eqf In order to simplify the
calculations let us assume $\omega_0=0$, i.e. the clock (2) starts
its motion with zero angular velocity. It is interesting to note
that a certain condition on $r$ must be fulfilled in order to
prevent that the speed attained by (2) overcomes the speed of
light $c$ before that a full revolution, at least, is completed.
Indeed, for $\omega=\alpha T$ and $\phi=\alpha T^2/2$, where
$\phi$ is the azimuthal angle of (2), the orbital period of (2)
measured by (1) is \eqi
\bar{T}=2\sqrt{\rp{\pi}{\alpha}};\lb{periodo}\eqf from the
condition that $v=r\alpha \bar{T}<c$ it follows\footnote{Note that
this implies that, over one revolution, the hypothesis of locality
is fulfilled because $r<\mathcal{L}=c/\omega(\bar{T})$
\ct{locality}, so that the use of \rfr{dilat} in getting
\rfr{tempo} is fully adequate.}\eqi
r<\rp{c}{2\sqrt{\pi\alpha}}.\lb{condiz}\eqf For $\omega_0=0$
\rfr{tempo} yields \eqia
\Delta\tau_I^{(2)}&=&\rp{1}{2}\left[\bar{T}\sqrt{1-\left(\rp{\alpha
r \bar{T}}{c}\right)^2}+\left(\rp{c}{\alpha r
}\right)\arcsin\left(\rp{\alpha r \bar{T}
}{c}\right)\right]=\nonumber\\
&=&\sqrt{\rp{\pi}{\alpha}}\left[\sqrt{1-\left(\rp{2r\sqrt{\pi\alpha}}{c}\right)^2}+\left(\rp{c}{2r\sqrt{\alpha
\pi}
}\right)\arcsin\left(\rp{2r\sqrt{\pi\alpha}}{c}\right)\right]\lb{supertempo}.\eqfa
It can be seen that $\Delta\tau_I^{(2)}$ {\it does depend on the
angular acceleration} $\alpha$. By using \rfr{condiz}, from
\rfr{supertempo} it can be shown that
$\Delta\tau_I^{(2)}<\Delta\tau_I^{(1)}$, i.e. the moving clock
measures a proper time interval shorter than that measured by the
rest clock at their reunion. Indeed, for $0<p\equiv
2r\sqrt{\pi\alpha}/c<1$, the condition
\eqi\sqrt{1-p^2}+\rp{\arcsin p}{p}<2,\nonumber\eqf to which
$\Delta\tau_I^{(2)}<\bar{T}$ reduces, is always satisfied. For
$\alpha=0$ and $\omega=$const, i.e. a circular uniform motion, it
is easy to show that  \eqi
\Delta\tau_I^{(2)}=\rp{2\pi}{\omega}\sqrt{1-\left(\rp{r\omega}{c}\right)^2}\equiv
\Delta\tau_I^{(1)}\sqrt{1-\left(\rp{r\omega}{c}\right)^2}<\Delta\tau_I^{(1)},\eqf
as can be found at pag.297 of \ct{moller}.
\section{The point of view of the accelerated observer}\lb{secdue}
In this Section we will mainly follow the approach exposed in
\ct{moller}.

Let us consider an accelerated frame $A$ which $rotates$ with
respect to $I$ at a variable angular speed $\omega(T)$ directed
along the $Z$ axis. In it we will use the cylindrical spatial-like
coordinates $\{x^1\equiv r,x^2\equiv\vartheta, x^3\equiv z \}$.
The plane $\{r,\vartheta\}$ coincides with the plane $\{X,Y\}$. We
will confine to it. The coordinate time $t$ will be chosen so that
$t=T$. Indeed, a standard clock at $P(r,\vartheta, 0)$ will reckon
a proper time $d\tau=dT\sqrt{1-[r\omega(T)/c]^2}$. As shown at
pag.256 of \ct{moller}, on a rotating disk it is convenient to use
coordinate clocks which tick at any place
$1/\sqrt{1-[r\omega(T)/c]^2}$ faster than the corresponding
standard clock; then, $t=T$. The relation between $I$ and $A$
will, then, be
\begin{equation}\left\{\begin{array}{lll}X=r\cos[\vartheta+\varphi(t)],\\\\
Y=r\sin[\vartheta+\varphi(t)],\\\\
Z=z,\\\\
T=t ,\lb{tras}\end{array}\right.\nonumber\end{equation}
Let us assume that $\varphi(t)=\varphi_0+\omega_0 t+\alpha t^2/2$,
i.e. a {\it uniformly accelerated circular} motion with
$\omega=\omega_0+\alpha t$ and constant angular acceleration
$\alpha$. So,
\begin{equation}\left\{\begin{array}{lll}
dX=dr\cos[\vartheta+\varphi(t)]-r\sin[\vartheta+\varphi(t)][(\omega_0+\alpha t)dt+d\vartheta]\\\\
dY=dr\sin[\vartheta+\varphi(t)]+r\cos[\vartheta+\varphi(t)][(\omega_0+\alpha t)dt+d\vartheta],\\\\
dZ=dz,\\\\
dT=dt ,\end{array}\right.\nonumber\end{equation} Then, the
spacetime metric is
\eqi(ds)^2=\left\{1-\left[\rp{r\omega(t)}{c}\right]^2\right\}(cdt)^2-(dr)^2-(dz)^2-
2\left[\rp{r\omega(t)}{c}\right]^2 cdtd\vartheta.\lb{spt}\eqf Let
us define \begin{equation}\left\{\begin{array}{lll}\chi\equiv
-\rp{[r\omega(t)]^2}{2},\\\\
\gamma_i\equiv-\rp{g_{i0}}{\sqrt{g_{00}}},\ i=1,2,3,\\\\
\gamma_{ij}\equiv-g_{ij}+\gamma_i\gamma_j,\ i,j=1,2,3,\\\\
v^i\equiv\rp{dx^i}{dt},\ i=1,2,3,\\\\
v\equiv\sqrt{\gamma_{ij}\rp{dx^i}{dt}\rp{dx^j}{dt}}.
\lb{definizioni}\end{array}\right.\eqf \Rfr{definizioni} shows
that the spacetime in a rotating frame is not time--orthogonal due
to the off--diagonal components of the metric tensor which induce
the so called vector potential ${\boldsymbol\gamma}$.
The general expression of the proper time interval of a particle
moving with respect to $A$ can be obtained as follows. From
\eqi(cd\tau)^2\equiv(ds)^2=g_{00}(cdt)^2+g_{ij}dx^i
dx^j+2g_{0k}cdtdx^k\nonumber\eqf it can be written
\eqi(d\tau)^2=(dt)^2\left(g_{00}+2\rp{g_{0k}}{c}\rp{dx^k}{dt}+\rp{g_{ij}}{c^2}\rp{dx^i}{dt}\rp{dx^j}{dt}\right).\lb{interm}\eqf
By adding and subtracting $[(\gamma_k/c)(dx^k/dt)]^2$ to
\rfr{interm} and by using \rfr{definizioni} it is possible to
obtain \eqi d\tau=dt\sqrt{\left(\sqrt{1+\rp{2\chi}{c^2}}-\rp{
\gamma_k v^k
}{c}\right)^2-\left(\rp{v}{c}\right)^2}.\label{deti}\eqf As shown
at pag. 279 of \ct{moller}, the components of the acceleration
experienced by a point at rest (i.e. for which $dx^i/d\tau=0$) in
the $\{r,\vartheta\}$ plane of $A$ can be obtained from the
geodesics equations. They are given by \eqi
a_i\equiv\gamma_{ij}\rp{d^2 x^j}{dt^2}=-\rp{\partial
\chi}{\partial x^i}-c\sqrt{1+\rp{2\chi}{c^2}}\rp{\partial \gamma_i
}{\partial t},\ i=1,2,3.\lb{accel}\eqf From \rfr{spt} and
\rfr{definizioni} we have
\begin{equation}\left\{\begin{array}{lll}
\gamma_2=\rp{r^2\omega(t)}{c\sqrt{1-\left[\rp{r\omega(t)}{c}\right]^2}},\ \gamma_1=\gamma_3=0,\\
\\
\gamma_{22}=r^2+\rp{r^4\omega^2(t)}{c^2\left\{1-\left[\rp{r\omega(t)}{c}\right]^2\right\}},\
\gamma_{11}=\gamma_{33}=1.
\end{array}\right.\lb{gammavari}\end{equation}
For a uniformly accelerated motion, i.e.
$\omega(t)=\omega_0+\alpha t$, \rfr{definizioni}, \rfr{accel} and
\rfr{gammavari} yield
\begin{equation}\left\{\begin{array}{lll}
a_{\rm centrifugal}\equiv a_r=r(\omega_0+\alpha
t)^2=r\omega^2(t),\\\\ a_{\rm
tangential}\equiv\rp{a_{\vartheta}}{r}=-\alpha r.
\end{array}\right.\lb{trasci}\end{equation}
They are just the Newtonian inertial dragging accelerations which
appear in an accelerated rotating frame
$-{\boldsymbol\omega}\times({\boldsymbol\omega}\times{\bm
r})-\dot{\boldsymbol\omega}\times{\bm r}$. By defining the spatial
distance between due nerby points as
$d\sigma\equiv\sqrt{\gamma_{ij}dx^i dx^j}$, from \rfr{gammavari}
it follows that, in general, the adopted frame is not rigid
because $d\sigma$ changes in time. However, it is not so for
points lying along a radius at constant $\vartheta$, as is the
case for (1) and (2). So, it makes sense to speak about orbits of
constant radius $r$.

Let us, now, consider the circular motion which (1) performs with
respect to (2) which is fixed in $A$. In this case the angular
velocity of (1) will be, of course, $-\omega(t)$. In order to find
$\Delta\tau_A^{(1)}$ we must use \rfr{deti}. \Rfr{gammavari} tells
us that
\begin{equation}\left\{\begin{array}{lll}
-\rp{\gamma_k
v^k}{c}=\rp{r^2\omega^2(t)}{c^2\sqrt{1-\left[\rp{r\omega(t)}{c}\right]^2}},\\\\
\left(\rp{v}{c}\right)^2 =
\rp{\gamma_{22}\omega^2(t)}{c^2}=\rp{r^2\omega^2(t)}{\cq}+\rp{r^4\omega^4(t)}
{c^2\left\{1-\left[\rp{r\omega(t)}{c}\right]^2\right\}}.
\end{array}\right.\lb{cazzivari}\end{equation}
By substituting \rfr{cazzivari} in \rfr{deti} shows that the
occurrence of the vector potential induces a cancellation which
yields $d\tau=dt=dT$, so that
\eqi\Delta\tau_A^{(1)}=\Delta\tau_I^{(1)}.\eqf

In regard to the proper time interval measured by (2), it can be
easily obtained by putting $v^i=v^2=0$ into \rfr{deti}. The result
is $d\tau=dT\sqrt{1-\left[r\omega(T)/c\right]^2}$, i.e.
\eqi\Delta\tau_A^{(2)}=\Delta\tau_I^{(2)}.\eqf

\section{Conclusions}
In this paper we have analyzed a particular version of the so
called clock paradox in which the moving clock, denoted as (2),
meets again the rest clock, denoted as (1), by performing an {\it
uniformly accelerated circular motion} of constant radius. The
inertial frame in which (1) is at rest is $I$ while the rotating
accelerated frame in which (2) is at rest is $A$. In particular,
we have investigated if the presence of a tangential acceleration
along the direction of motion of (2) is responsible for a
differential aging between the two clocks at their reunion.

The obtained results can be summarized as follows
\begin{itemize}
  \item  Both for $\omega=$const and for $\omega$ linearly varying with
  time,
  $\Delta\tau_I=\Delta\tau_A$ for any given clock, as expected.
  So, the Special and General Theories of Relativity do not lead
  to inconsistencies.
  \item For $\omega=$const the differential aging is given by the
  usual Special Relativistic formula, based on the hypothesis of
  locality,
  $\Delta\tau^{(2)}/\Delta\tau^{(1)}=\sqrt{1-\left({v}/{c}\right)^2}<1$ with $v=r\omega$
  \item The magnitude of the proper time intervals {\it does depend on the angular acceleration $\alpha$}
  for an {\it uniformly accelerated circular} motion. Moreover,
  for $\omega$ linearly varying in time it turns out that
  $\Delta\tau^{(2)}<\Delta\tau^{(1)}$.
  \item The differential aging does depend on $\alpha$, i.e.
  $\Delta\tau^{(2)}/\Delta\tau^{(1)}=f(\alpha)$,
  contrary to the case in which an accelerated rectilinear motion is
  considered in which $\Delta\tau^{(2)}/\Delta\tau^{(1)}$
  depends only on the velocity reached when the force is inverted.
\end{itemize}

\end{document}